\documentclass[aps,prl,reprint,a4paper,showkeys,showpacs,superscriptaddress]{revtex4-1}

\usepackage{graphicx}

\usepackage{amssymb}
\usepackage{amsmath}
\usepackage{amsfonts}
\begin{document}

\title{Soft-Hair-Enhanced Entanglement Beyond Page Curves\\
in a Black Hole Evaporation Qubit Model}

\begin{abstract}
We propose a model with multiple qubits that reproduces the thermal
properties of four-dimensional Schwarzschild black holes (BHs) by
simultaneously taking account of the emission of Hawking particles and the
zero-energy soft-hair evaporation at horizon. The results verify that the
entanglement entropy between a qubit and other subsystems, including emitted
radiation, is much larger than the BH entropy analogue of the qubit, as
opposed to the Page curve prediction. Our result suggests that early Hawking
radiation is entangled with soft hair, and that late Hawking radiation can
be highly entangled with the degrees of freedom of BH, avoiding the
emergence of a firewall at the horizon.
\end{abstract}

\author{Masahiro Hotta}
\affiliation{Graduate School of Science, Tohoku University,\\ Sendai, 980-8578, Japan}
\author{Yasusada Nambu}
\affiliation{Graduate School of Science, Nagoya University,\\ Nagoya, 464-8601, Japan}
\author{Koji Yamaguchi}
\affiliation{Graduate School of Science, Tohoku University,\\ Sendai, 980-8578, Japan}
\maketitle

\ \ \textit{Introduction}.---There has been a rapid increase in the
importance of the entanglement entropy (EE) of macroscopic systems in
quantum gravity and condensed matter physics. After the advent of the
Ryu-Takayanagi formula \cite{RT}, which shows that the EE in a $D$-dimensional
($D$-dim) conformal field theory is equal to the Bekenstein-Hawking entropy
in a $(D+1)$-dim gravity theory with a negative cosmological constant, the EE
sheds light on unexpected features of spacetimes. The formula indicates the
interesting possibility that quantum information generates curved spacetime
in a holographic way \cite{J}. In condensed matter physics, EE plays the
role of an exotic order parameter for topological insulators \cite{KP,LW}. Recently, the EE was directly measured in an ultracold bosonic atom
experiment \cite{exp}, which indicated that the EE time evolution of
many-body systems may also be observed experimentally. In the theory , the
first-principles calculation of EE evolution is very complicated for
macroscopic systems, and it has not been achieved to date.

In discussions related to black hole (BH) physics, a famous conjecture of EE
evolution, i.e., the Page curve, is often adopted. Hawking \cite{h} showed
that BHs evaporate by emitting thermal Hawking radiation. It is possible to
assume a thought experiment in which the initial state is a pure state and
evolves in a unitary way. AdS/CFT arguments support the unitarity of the
process \cite{h2}. Page conjectured an EE evolution between an evaporating
BH and its radiation \cite{page}, and the evolution is called a Page curve.
It is assumed that the EE is equal to the BH thermal entropy after the
so-called Page time. Though the conjecture is based on analyses of general
many-body systems without horizons, typicality arguments of statistical
mechanics are expected to provide essentially the same behavior in various
physical systems including BHs.

In this Letter, we argue a possibility that EE of a BH subsystem shared with
other systems including radiation is much larger than its BH thermal entropy
after the Page time, using a thermal system of decaying $N$ qubits. This
implies the direct breakdown of Page's conjecture. The key idea is to
reproduce the Hawking temperature relation of 4-dim Schwartzschild BHs in
the qubit system:%
\begin{equation}
T_{\mathrm{rad}}=\frac{1}{8\pi G\left\langle H_{N}\right\rangle },  \label{r1}
\end{equation}%
where $T_{\mathrm{rad}}$ is the temperature of thermal radiation emitted by the
system, $\left\langle H_{N}\right\rangle $ is the quantum expectation value of
the system Hamiltonian, and $G$ is the gravitational constant. Natural units are
adopted ($c=\hbar =k_{B}=1$). In the same way as in BH physics, we are able
to introduce a Bekenstein-Hawking entropy $S_{\mathrm{BH}}$ by defining the system
temperature $T$ equal to $T_{\mathrm{rad}}$, and integrating the first law; $%
dS_{\mathrm{BH}}=d\left\langle H_{N}\right\rangle /T=8\pi G\left\langle
H_{N}\right\rangle d\left\langle H_{N}\right\rangle $. Then $S_{\mathrm{BH}}$ takes
the same value of the area-law entropy given by $4\pi r^{2}/(4G)$ with
radius $r=2G\left\langle H_{N}\right\rangle $: 
\begin{equation}
S_{\mathrm{BH}}=\left( 16\pi GT^{2}\right) ^{-1}.  \label{r2}
\end{equation}%
The EE and the thermal entropy are compared with $S_{\mathrm{BH}}$ in this model.

It is worth stressing that the relation in Eq.~(\ref{r1}) yields a negative
heat capacity, which cannot appear in ordinary systems. In order to
incorporate this exotic aspect in our model, we consider a transition from a
qubit in a zero-energy state $|-\rangle $\ to an escaping zero-energy
particle of a field $\Psi _{S}$ which mimics soft hair in BH physics. The BH
soft-hair conjecture asserts that zero-energy degrees of freedom emerge at a
BH horizon \cite{HSS} \cite{HPS}. The soft-hair microstates contribute to
the BH entropy \cite{HSS} \cite{HTY}. Quantum information of infalling
matter behind the horizon is (at least partially) stored in the soft hair by
use of conserved Noether charges of would-be diffeomorphism \cite{HPS}.
While the horizon soft hairs evaporate, the quantum information may be
transmitted into Bondi-Metzner-Sachs (BMS) soft hairs at future null
infinity \cite{HPS} \cite{HPS2}. \ In our model, after emission of a
zero-energy $\Psi _{S}$ particle, the original system settles down in the
vacuum state $|0\rangle $ with zero energy, which stands for a lack of
qubit. The sum of the number of surviving qubits and $\Psi _{S}$ particles
is conserved. Thus the system in $|0\rangle $ never returns spontaneously
into $|-\rangle $ or $|+\rangle $ without ingoing $\Psi _{S}$ particles.
Since each qubit decay does not decrease the system energy, a smaller number
of qubits in the system carries the same energy. After a fast scrambling for
themalization of surviving qubits in a state of a sub-Hilbert space spanned
by $|\pm \rangle $, the system temperature $T$ becomes higher. This indeed
demonstrates the negative heat capacity behavior. A similar mechanism of
negative heat capacity generation was proposed in a D0-brane matrix model 
\cite{hanada}. There are earlier studies regarding the EE evolution of qubit
models \cite{M,G,A}; however, the negative heat capacity was not discussed.

In our model, quantum states of a decaying qubit belong to a 3-dim Hilbert
space spanned by $|\pm \rangle $ and $|0\rangle $, and thermalization occurs
only in its 2-dim $|\pm \rangle $ sector. This enables the system to possess
a larger EE than thermal entropy of the surviving qubit.

\bigskip

\bigskip

\begin{figure}[tbp]
\includegraphics[width=7.5cm]{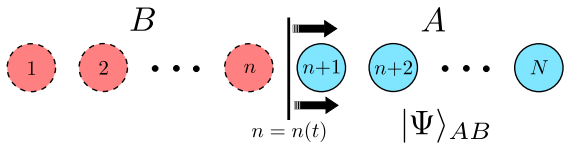}
\caption{Schematic diagram for the time evolution of decaying qubits in the
Page curve conjecture. The vertical line in this figure represents the
boundary that separates the subsystems.}
\label{fig1}
\end{figure}

\textit{Page curve: Summary and problems.}---For a pure state $|\Phi \rangle 
$ of a composite system $AB$, the bipartite entanglement between the
subsystems $A$ and $B$ is quantified by EE: $S_{\mathrm{EE}}=-\mathrm{Tr}_{A}\left[
\rho _{A}\ln {\rho _{A}}\right] $, where $\rho _{A}=\mathrm{Tr}_{B}\left[
|\Phi \rangle \langle \Phi |\right] $ is the reduced state of subsystem $A$.
The Page curve conjecture is based on a typicality theorem \cite{L,Seth,page}%
. The theorem implies that the EE values between two finite macroscopic
systems $A$ and $B$ in typical states of Hilbert space are very close to its
maximum value when $D_{A}\ll D_{B}$, where $D_{A}$ and $D_{B}$ denote the
dimensions of the sub-Hilbert space of $A$ and $B$, respectively. This state
typicality is defined using the Haar measure in the Hilbert space. If we
have a nontrivial Hamiltonian with a small interaction between $A$ and $B$,
then the state becomes a Gibbs state $\rho _{A}\left( T\right) $ at a finite
temperature $T$, with respect to the energy conservation of the total system
due to the Sugita theorem \cite{sugita,sugita2}. Thus, for typical pure
states of an $AB$ system, the EE between $A$\ and $B$ is very close to the
thermal entropy of $A$, which is computed from $\rho _{A}\left( T\right) $.
Let us consider $N$ qubits and put them on a line. At the initial time, all
the qubits are assigned to $A$ components. The total system is in a typical
pure state $|\Psi \rangle _{AB}$, which is randomly chosen from the Hilbert
space respecting energy conservation. As depicted in Fig.~\ref{fig1},
suppose a boundary separates the system into two parts, and moves from left
to right. 
In the Page curve conjecture, the decay process of $A\rightarrow B$ is
described by assigning the left-hand-side qubits to $B$ components. Plotting
the thermal entropy of the smaller subsystem at each time generates a Page
curve. The conjecture asserts that this plot is the EE time evolution for
the decay process of $A\rightarrow B$ in high precision. The time at which
the thermal entropy of $A$ equates to the thermal entropy of $B$ is referred
to as the Page time, which corresponds to $n=N/2$. After the Page time, the
subsystems of $A$ are maximally entangled with $B$. Because of the entanglement
monogamy, the entanglement among the subsystems vanishes in the conjecture.

First, it should be noted that the Page curves have no dependence on the
decay dynamics. This appears unusual, since details of realistic decay
channels can affect the EE evolution, and the conjecture is not capable of
discriminating the following two EE evolutions. In the first case (i), fast
scrambling interaction in the thermal equilibrium of $A$ components is
switched off before the decay of $A$. After that, each $A$ component decays
independently into a $B$ component, and no thermalization interaction
appears during the process. In the second case (ii), fast scrambling to the
equilibrium of $A$ components continues during the decay of $A\rightarrow B$%
. In BH physics, fast scrambling for an evaporating BH has been proposed 
\cite{fs} and is supposed to justify the random selection of pure states of
the BH and its radiation. However, the detailed properties remain elusive.

Another weak point of the conjecture is that this is based on many-body
systems with positive heat capacity. Actually, we have no plausible arguments
for negative heat capacity cases like BH evaporation so far.

\bigskip

\textit{Qubit} \textit{model.}---Let us introduce decaying $N$ qubits by
simultaneously emitting zero-energy radiation of soft hair and energetic
radiation corresponding to Hawking radiation. The free Hamiltonian of a
decaying single qubit is given by 
\begin{equation}
H=\omega |+\rangle \langle +|,  \label{r0}
\end{equation}%
where $\omega $ is a positive constant and $|+\rangle $, $|-\rangle $,
and $|0\rangle $ are the eigenstates of the energy eigenvalues $\omega $, $0$
and $0$, respectively. We have a fast scrambling interaction $H_{\mathrm{fs}}$ for
surviving qubits. The scrambling process is assumed to be much faster than
the emission of particles. It occurs only among qubits in the $|\pm \rangle $
sector state, which preserves the total energy, and does not affect
subsystems in $|0\rangle $. A simple two-body interaction example of $H_{\mathrm{fs}}$
is given by 
\begin{align*}
H_{\mathrm{fs}}& =\sum_{s_{1}=-}^{+}\cdots
\sum_{s_{N}=-}^{+}\sum_{i=1}^{N}\sum_{j=i+1}^{N}h\left( s_{i}s_{j}\right) \\
& \times I_{1}\otimes \cdots I_{i-1}\otimes |s_{i}\rangle \langle
s_{j}|\otimes I_{i+1}\cdots \\
& \otimes I_{j-1}\otimes |s_{j}\rangle \langle s_{i}|\otimes I_{j+1}\otimes
\cdots I_{N},
\end{align*}%
where $h\left( s_{i}s_{j}\right) $ are real coupling constants and $I_{i}$
is the unit matrix for the $i$th site subsystem. At $t=-t_{o}<0$, the $N$
qubits are set in a pure state, which is composed of $|\pm \rangle $, with
the total energy expectation value fixed as $E_{i}$. By acting the scrambling
operator $\exp (-it_{o}H_{\mathrm{fs}})$ onto the initial pure state, we get a
typical pure state at $t=0$. In a typical pure state, each qubit is in a
Gibbs state $\rho (T(0))$ at temperature $T(0)$. Throughout this Letter, $%
\rho (T)$ represents the Gibbs state of the two-level system at temperature $%
T$, i.e., $\rho (T)=\left( |-\rangle \langle -|+\exp {(-\omega /T)}|+\rangle
\langle +|\right) /(1+\exp {(-\omega /T)})$. $T(0)$ is uniquely determined
by $E_{i}$ via $T(0)=\omega /\ln \left( \frac{N\omega }{E_{i}}-1\right) $.

In our model, a Hawking particle with energy $\omega $ is emitted out of the
qubit flipping $|+\rangle $ to $|-\rangle $. The particle escapes from the
qubit along a real axis denoted by $x$ and propagates to $x=+\infty $. The
state of one particle at some position $x$ is described using a 1-dim
bosonic Schr\"{o}dinger field $\Psi _{R}(x)$, as $\Psi _{R}^{\dag
}(x)|\mathrm{vac}\rangle $. Here, $|\mathrm{vac}\rangle $ is the vacuum state of the field
that satisfies $\Psi _{R}(x)|\mathrm{vac}\rangle =0$, and the initial state of the
field. The effective Hamiltonian that describes the emission is given by 
\begin{align}
H_{R}& =\int \Psi _{R}^{\dag }(x)\left( -i\partial _{x}\right) \Psi
_{R}(x)dx+|-\rangle \langle +|\int g(x)\Psi _{R}^{\dag }(x)dx  \notag \\
& \qquad \qquad \qquad +|+\rangle \langle -|\int g^{\ast }(x)\Psi _{R}(x)dx.
\label{-2}
\end{align}%
$g(x)$ is a localized function around the qubit and provides coupling
between the qubit and $\Psi _{R}$. The unitary time-evolution operator is
given by $U_{R}=\exp \left( -itH_{R}\right) $. The dynamics conserves the
excitation number, $|+\rangle \langle +|+\int \Psi _{R}^{\dag }(x)\Psi
_{R}(x)dx$. Therefore, when a qubit is in $|+\rangle $, the composite system
evolves into an entangled state, such that 
\begin{equation}
U_{R}|+\rangle |\mathrm{vac}\rangle =c_{+}(t)|+\rangle |\mathrm{vac}\rangle +|-\rangle \left(
\int \varphi (x,t)\Psi _{R}^{\dag }(x)dx\right) |\mathrm{vac}\rangle ,  \label{-1}
\end{equation}%
where $\varphi (x,t)$ is the wave function of the created particle. The
survival probability of $|+\rangle $ \ is given by $\left\vert
c_{+}(t)\right\vert ^{2}$. This dynamics can be solved, as shown in
Supplemental Material \cite{SM}. A quantum channel for the qubit is
introduced as 
\begin{equation}
\Gamma _{R}\left[ \rho \right] =\mathrm{Tr}_{\Psi _{R}}\left[ U_{R}\left(
\rho \otimes |\mathrm{vac}\rangle \langle \mathrm{vac}|\right) U_{R}^{\dag }\right] ,
\label{-3}
\end{equation}%
and the evolution of the qubit in $|+\rangle\langle+|$ is given by 
\begin{equation}
\Gamma _{R}\left[ |+\rangle \langle +|\right] =\left( 1-r\right) |+\rangle
\langle +|+r|-\rangle \langle -|,
\end{equation}%
where $r$ is the probability of finding $|-\rangle $ and is given by $%
r=1-\left\vert c_{+}(t)\right\vert ^{2}$. The quantum channel satisfies $%
\Gamma _{R}\left[ |0\rangle \langle 0|\right] =|0\rangle \langle 0|$ and $%
\Gamma _{R}\left[ |-\rangle \langle -|\right] =|-\rangle \langle -|$ due to
the conservation of the excitation number. Although the field quanta created by
the transition trigger reexcitation from $|-\rangle $ into $|+\rangle $
while $\varphi (x,t)$ has nonzero overlap with $g(x)$, this is not
essential. Indeed, the probability $r~$monotonically decreases with time due
to the leakage of $\varphi $ out of the overlap region. The transition from $%
|-\rangle $ into $|+\rangle $ occurs for a qubit by the fast scrambling with
other qubits.

Now let us introduce another channel for the decay of a zero-energy qubit into a
zero-energy soft-hair particle. Suppose a similar Hamiltonian for the
process: 
\begin{align}
H_{S}& =\int \Psi _{S}^{\dag }(x)\left( -i\partial _{x}\right) \Psi
_{S}(x)dx+|0\rangle \langle -|\int g(x)\Psi _{S}^{\dag }(x)dx  \notag \\
& \qquad \qquad \qquad +|-\rangle \langle 0|\int g^{\ast }(x)\Psi _{S}(x)dx,
\label{r3}
\end{align}%
where $\Psi _{S}(x)$ mimics the soft hair and satisfies $\Psi _{S}(x)$ $%
|\mathrm{vac}\rangle =0$. We then obtain a quantum channel for the decay: 
\begin{equation*}
\Gamma _{S}\left[ \rho \right] =\mathrm{Tr}_{\Psi _{S}}\left[ U_{S}\left(
\rho \otimes |\mathrm{vac}\rangle \langle \mathrm{vac}|\right) U_{S}^{\dag }\right] ,
\end{equation*}%
where $U_{S}=\exp \left( -itH_{S}\right) $. This satisfies $\Gamma _{S}\left[
|+\rangle \langle +|\right] =|+\rangle \langle +|$, while a qubit in $%
|-\rangle $ decays into $\Psi _{S}$ with probability $s$: $\Gamma _{S}\left[
|-\rangle \langle -|\right] =(1-s)|-\rangle \langle -|+s|0\rangle \langle 0|$%
. Once the decay occurs, the qubit cannot come back by the next operation of 
$\Gamma _{S}$, i.e., $\Gamma _{S}\left[ |0\rangle \langle 0|\right]
=|0\rangle \langle 0|$. Also the fast scrambling does not cause the
transition from $|0\rangle $ to $|\pm \rangle $. 

\begin{figure}[tbp]
\includegraphics[width=7.5cm]{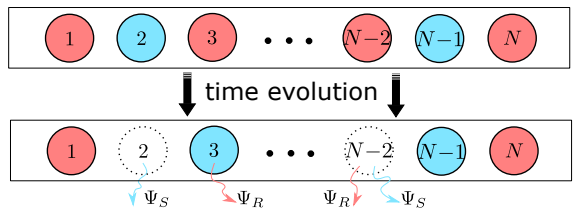}
\caption{Schematic diagram of the decaying qubit model with negative heat
capacity. When a qubit makes a transition from $|+\rangle\langle +|$ (red
circle) into $|-\rangle\langle -|$ (blue circle), a $\Psi_R$ particle is
emitted. After the decay of qubits into $\Psi_S$, there remains $%
|0\rangle\langle 0|$ (dotted circle).}
\label{fig2}
\end{figure}

By combining $\Gamma _{S}$ and $\Gamma _{R}$, we define a channel $\Gamma
=\Gamma _{S}\Gamma _{R}$. A schematic diagram of this model is given in Fig.~%
\ref{fig2}.

A qubit in a Gibbs state at initial temperature $T(0)$ evolves by $\Gamma $
as 
\begin{equation}
\Gamma \left[ \rho \left( T(0)\right) \right] =\left( 1-p\right) |0\rangle
\langle 0|+p\rho \left( T\right) ,  \label{100}
\end{equation}%
where the survival probability $p$ is given by 
\begin{equation*}
p=1-s\left( 1+r\exp \left( -\omega /T(0)\right) \right) /\left( 1+\exp
\left( -\omega /T(0)\right) \right) ,
\end{equation*}%
and $T=\omega /\ln \left[ \frac{1-s}{1-r}\left( \exp \left( \omega
/T(0)\right) +r\right) \right] $. Although the state in Eq.~(\ref{100}) is
not a Gibbs state of the 3-dim Hilbert space, we are able to identify $T$ as
the temperature of the decaying qubit system, because the thermal flux of $%
\Psi _{R}$ divided by the number of surviving qubits is determined by $T$.
Let us impose a natural condition that $T$ is infinity when $T(0)$ is
infinity. This fixes a nontrivial relation between the two probability
parameters $r$ and $s$ as $s=2r/(1+r)$. First, let us consider case (i). The
temperature $T$ rises when the radiation emission probability $r$ increases
as 
\begin{equation}
T=\frac{\omega }{\ln \left[ \frac{\exp \left( \omega /T(0)\right) +r}{1+r}%
\right] },  \label{50}
\end{equation}%
although the particle extracts energy from the system. This implies that the
model realizes a negative heat capacity. However, the temperature does not
go to infinity at $r=1$. This feature is different from that of BH
evaporation, where the BH temperature goes to infinity at the last burst.

\textit{Continuous scrambling.}--- On the other hand, continuous fast
scrambling (case (ii)) makes the final temperature infinity. As shown in Supplemental Material \cite{SM}, the one-qubit Gibbs state remains unchanged after each fast
scrambling and qubit free evolution. Every fast scrambling loses the
correlation between a single qubit and the fields. Let us consider $t$ times
operation of $\Gamma $ and take the $r\rightarrow 0$ limit with $rt=\tau $
fixed, $\Gamma ^{t}\left[ \rho \left( T(0)\right) \right] =\left( 1-p(\tau
)\right) |0\rangle \langle 0|+p(\tau )\rho \left( T(\tau )\right) $. This
provides the dynamics of a single decaying qubit reduced from the full
dynamics of $N$ decaying qubits, $\Psi _{R}$ and $\Psi _{S}$ with the total
Hamiltonian:

\begin{equation}
H_{tot}=\sum_{i=1}^{N}\left( H_{i}+H_{R_{i}}+H_{S_{i}}\right) +H_{\mathrm{fs}},
\label{r5}
\end{equation}%
where $H_{i}$ is the $i$ th subsystem free Hamiltonian of Eq.~(\ref{r0}), $H_{R_{i}}$ and $H_{S_{i}}$ are the $i$ th field Hamiltonians of Eqs.~(%
\ref{-2}) and ~(\ref{r3}), and $H_{\mathrm{fs}}$ is the fast scrambling
Hamiltonian for the ($|+\rangle$, $|-\rangle$) sector. Here, as usual, we
assume that the Hamiltonian of evaporating qubits can be approximated by the
free Hamiltonian contribution: $H_{N}=\sum_{i}H_{i}$.

In order to reproduce the BH thermal properties in this model, it is noted
that $N$ depends on $T(0)$, and $N=\left( 4\pi G\omega T(0)\right) ^{-1}$
should hold. The energy $\omega $ is also assumed to be much smaller than $%
T(0)$. The constraint $\omega \ll T(0)$ yields 
\begin{equation}
p(\tau )\approx\exp (-\tau ),  \label{51}
\end{equation}%
and since the temperature is given by 
\begin{equation}
T(\tau )\approx T(0)\exp \tau ,  \label{52}
\end{equation}%
the temperature at $\tau =\infty $ is infinity. Eqs.~(\ref{50}), (\ref{51})
and (\ref{52}) are derived in Supplemental Material \cite{SM}. One qubit energy $E=\mathrm{%
Tr}\left[ H\rho \right] $ is computed as 
\begin{equation*}
E(\tau )=p(\tau )\frac{\omega \exp \left( -\omega /T(\tau )\right) }{1+\exp
\left( -\omega /T(\tau )\right) }\approx p(\tau )\frac{\omega }{2}.
\end{equation*}%
Therefore, we have the relation $E=\omega T(0)/(2T)$ for a single qubit. The
expectation value of the total energy~$\left\langle H_{N}\right\rangle $ of $%
N$ qubits is given by 
\begin{equation*}
\left\langle H_{N}\right\rangle =NE=\left( 8\pi GT\right) ^{-1},
\end{equation*}%
and this is the precise relation in Eq.~(\ref{r1}) because $T_{\mathrm{rad}}=T$. Thus
the Bekenstein-Hawking entropy $S_{\mathrm{BH}}$ is defined by Eq.~(\ref{r2}) in this
model. By introducing a new time coordinate $t^{\prime }$, such that $\tau
=\ln {(1-t^{\prime }/t_{\mathrm{BH}})^{-1/3}}$, the qubit system decays completely at
finite lifetime $t^{\prime }=t_{\mathrm{BH}}$, just as 4-dim Schwarzschild BHs do.
The temperature at the Page time is evaluated as $T_{\mathrm{Page}}\approx 2T(0)$.

In the Page curve conjecture, one decaying qubit is almost maximally
entangled with emitted matter after the Page time and has no correlation
with other qubits due to entanglement monogamy. Therefore, $S_{\mathrm{EE}}$ between
one decaying qubit and other subsystems (other $N-1$ qubits+$\Psi _{R}$ +$%
\Psi _{S}$) must be equal to $S_{\mathrm{BH}}/N$ after the Page time. $S_{\mathrm{EE}}$ at
time $\tau $ is computed as%
\begin{equation*}
S_{EE}=-\mathrm{Tr}\left[ 
\begin{array}{c}
\left( \left( 1-p(\tau )\right) |0\rangle \langle 0|+p(\tau )\rho \left(
T(\tau )\right) \right) \\ 
\times \ln \left( \left( 1-p(\tau )\right) |0\rangle \langle 0|+p(\tau )\rho
\left( T(\tau )\right) \right)%
\end{array}%
\right] .
\end{equation*}%
As shown in Supplemental Material \cite{SM}, the average thermal entropy of the total system is
computed as $NpS_{\mathrm{th}}$, where $S_{\mathrm{th}}$ is the qubit thermal entropy at
temperature $T$, which is given by $S_{\mathrm{th}}=\ln \left( 1+\exp \left( -\omega
/T\right) \right) +\omega /(T\exp \left( \omega /T\right) +T)$. Thus, the
average of one-qubit thermal entropy is defined as $pS_{\mathrm{th}}$. In Fig.~\ref%
{fig3}, these three entropies are plotted as a function of $1-T(0)/T$. At
the initial time ($T=T(0)$), $S_{\mathrm{EE}}$ of one qubit does not vanish, because
the qubit is entangled with other qubits, although $S_{\mathrm{EE}}$ of $N$%
 qubits vanishes at that time. The entropies are conjectured to be equal to
each other after the Page time. Actually, they behave very differently ($%
S_{\mathrm{BH}}/N\ll pS_{\mathrm{th}}\ll S_{\mathrm{EE}}$). In a high-temperature regime with $T\gg T(0)
$, they are analytically evaluated as 
\begin{align*}
S_{\mathrm{BH}}/N& \sim \frac{\omega }{4T(0)}\left( \frac{T(0)}{T}\right) ^{2}, \\
pS_{\mathrm{th}}& \sim \frac{T(0)}{T}\ln 2, \\
S_{\mathrm{EE}}& \sim \frac{T(0)}{T}\ln \left( \frac{T}{T(0)}\right) .
\end{align*}%
The difference between $S_{\mathrm{BH}}/N$ and $pS_{\mathrm{th}}$ originates from the extra
factor $p$ in the energy $E=pE_{\mathrm{th}}$. The discrepancy is derived in Supplemental Material \cite{SM}. The difference between $pS_{\mathrm{th}}$ and $S_{\mathrm{EE}}$ is the binary
entropy of $p$ that reflects the non-trivial contribution of the vacuum
states, $|0\rangle $. Thus, in this model, the Page curve conjecture does
not work.

\begin{figure}[tbp]
\includegraphics[width=7.5cm]{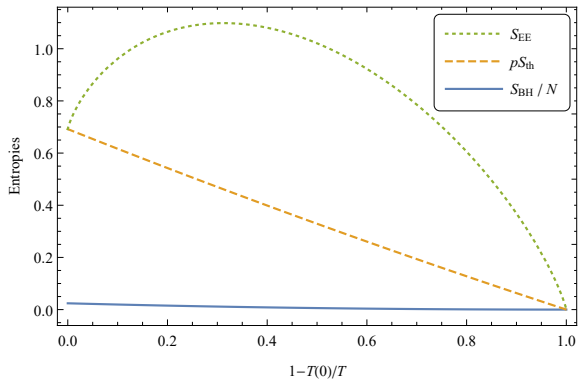}
\caption{Plots of three types of entropies in the decaying qubit model at $%
\protect\omega/T(0)=0.1$. The Page time is defined by $p=1/2$, and
corresponds to $T(0)/T_{\mathrm{Page}}\approx 0.52$.}
\label{fig3}
\end{figure}

\textit{Conclusion and discussion. }--- In a model of decaying qubits into
zero-energy degrees of freedom, the thermal properties of 4-dim
Schwarzschild BH evaporation are precisely reproduced. The EE is much larger
than the average thermal entropy and the Bekenstein-Hawking entropy analogue
for each qubit. This is the first result of a breakdown of the Page curve
ansatz in a model which satisfies the Hawking temperature relation in Eq.~(%
\ref{r1}).

The result provides a new feature for a resolution of the information loss
problem. In our model, the emission of a Hawking particle of $\Psi _{R}$ at an
early stage makes a transition from $|+\rangle $ into $|-\rangle $ of a
qubit. After the Page time, the qubit in $|-\rangle $ almost decays into a
zero-energy particle of $\Psi _{S}$, which may be interpreted as BMS soft
hair propagating to future null infinity. This suggests that in the BH
firewall paradox \cite{AMPS,B,HS}, early Hawking radiation is entangled with
zero-energy BMS soft hair and that late Hawking radiation can be highly
entangled with the degrees of freedom of the BH (surviving qubits in $%
|-\rangle $), avoiding the emergence of a firewall at the horizon.

The soft hair influence for black holes with positive heat capacity like
large AdS black holes remains elusive. 

\bigskip

\textit{Acknowlegements}. ---The authors thank Masanori Hanada
and Hal Tasaki for useful discussions. This research was partially supported
by Kakenhi Grants-in-Aid No. 16K05311 (M.H.) and No. 16H01094 (Y.N.) from the
Japan Society for the Promotion of Science (JSPS) and by the Tohoku
University Graduate Program on Physics for the Universe (K.Y.).

\clearpage
\widetext
\appendix
\section{SUPPLEMENTAL MATERIAL}

\makeatletter
\renewcommand{\theequation}{    S\arabic{equation}} %
\@addtoreset{equation}{section} \makeatother

\begin{center}
{\large Decay Channel }
\end{center}

Substituting Eq.~(5) into the Schr\"{o}dinger equation with Hamiltonian
Eq.~(4) yields the following equations of motion:%
\begin{equation}
i\frac{d}{dt}c_{+}(t)=\int g^{\ast }(x)\varphi (x,t)dx,  \label{s1}
\end{equation}

\begin{equation}
\left( \partial_{t}+\partial_{x}\right) \varphi(x,t)=-ic_{+}(t)g(x).
\label{s2}
\end{equation}
Using $x^{\pm}=x\pm t$, Eq.~(\ref{s2})\bigskip\ is rewritten as 
\begin{equation*}
\partial_{+}\varphi=-\frac{i}{2}c_{+}\left( \frac{1}{2}\left(
x^{+}-x^{-}\right) \right) g\left( \frac{1}{2}\left( x^{+}+x^{-}\right)
\right) .
\end{equation*}
Taking account of the retarded boundary condition, integration of the above
equation is achieved as 
\begin{equation}
\varphi=-\frac{i}{2}\int_{x^{-}}^{x^{+}}c_{+}\left( \frac{1}{2}\left(
x^{\prime}-x^{-}\right) \right) g\left( \frac{1}{2}\left(
x^{\prime}+x^{-}\right) \right) dx^{\prime}.  \label{s3}
\end{equation}
Substituting Eq.~(\ref{s3}) into Eq.~(\ref{s1}) yields

\begin{equation*}
i\frac{d}{dt}c_{+}(t)=-\frac{i}{2}\int dxg^{\ast }(x)\int_{x-t}^{x+t}g\left( 
\frac{1}{2}\left( x^{\prime }+x-t\right) \right) c_{+}\left( \frac{1}{2}%
\left( x^{\prime }-x+t\right) \right) dx^{\prime }.
\end{equation*}%
By changing the integral variable from $x^{\prime }$ to $t^{\prime }$, such
that 
\begin{equation*}
t^{\prime }=\frac{1}{2}\left( x^{\prime }-x+t\right) ,
\end{equation*}%
the following equation is derived.

\begin{equation*}
\frac{d}{dt}c_{+}(t)=-\int_{0}^{t}K(t-t^{\prime })c_{+}(t^{\prime
})dt^{\prime },
\end{equation*}%
where $K(t)=\int g^{\ast }(x)g\left( x-t\right) dx$. This is generally
solved using the Laplace transformation. As a simple example, let us take $%
g(x)=\lambda \delta \left( x\right) $ with positive $\lambda $. The survival
probability of the up state is computed as $\left\vert c_{+}(t)\right\vert
^{2}=\exp \left( -\lambda ^{2}t\right) $. The wave function $\varphi (x,t)$
is given by%
\begin{equation*}
\varphi (x,t)=-i\lambda \exp \left( -\frac{\lambda ^{2}}{2}(t-x)\right)
\Theta \left( t-x\right) \Theta (x).
\end{equation*}%
If we take a large $\lambda $, then we have a localized wave packet out of a
qubit.

The feature of zero energy states comes from gravitational physics. The
zero-enery qubit decay happens due to a large phase space volume for the
soft hairs. In ordinally physical systems, the phase space is narrow and the
zero-energy decay can be omitted.

\bigskip

\begin{center}
\bigskip{\large Derivation of Temperature Evolution }
\end{center}

In this section, we derive Eqs. (10), (12), and (13). By using $\Gamma _{R}$
and $\Gamma _{S}$, we obtain the following result: 
\begin{align*}
& \Gamma\left[ \rho\left( T\right) \right] \\
& =s\frac{1+r\exp\left( -\omega/T\right) }{1+\exp\left( -\omega/T\right) }%
|0\rangle\langle0|+(1-s)\frac{1+r\exp\left( -\omega/T\right) }{1+\exp\left(
-\omega/T\right) }|-\rangle\langle-|+(1-r)\frac{\exp\left( -\omega/T\right) 
}{1+\exp\left( -\omega/T\right) }|+\rangle\langle+|.
\end{align*}
The state is then expressed as 
\begin{equation*}
\Gamma\left[ \rho\left( T\right) \right] =\left( 1-p\right)
|0\rangle\langle0|+p\rho\left( T^{\prime}\right) ,
\end{equation*}
where the temperature $T^{\prime}$ after the operation is given by 
\begin{equation}
T^{\prime }=\frac{T}{1+\frac{T}{\omega }\ln \left( 1+r\exp \left( -\omega
/T\right) \right) -\frac{T}{\omega }\ln \frac{1-r}{1-s}}.  \label{s4}
\end{equation}%
and the probability $p$ is given by 
\begin{equation*}
p=\left( 1-r\right) \frac{\exp \left( \omega /T^{\prime }\right) +1}{\exp
\left( \omega /T\right) +1}.
\end{equation*}%
Imposing $1/T=1/T^{\prime }=0$ in Eq.~(\ref{s4}) yields 
\begin{equation}
s=\frac{2r}{1+r}.  \label{s5}
\end{equation}%
Substituting Eq.~(\ref{s5}) into Eq.~(\ref{s4}) and replacing $T\rightarrow
T(0)\,\ $and $T^{\prime }\rightarrow T$ provides Eq.~(10). \ By performing $%
t $ times $\Gamma ,$ the temperature after the operation is computed as 
\begin{equation}
T=\frac{\omega }{\ln \left[ 1+\left( \exp \left( \frac{\omega }{T(0)}\right)
-1\right) \left( 1+r\right) ^{-t}\right] }.  \label{s6}
\end{equation}%
Similarly, the survival probability of a qubit after the operation is given
by%
\begin{equation}
p=\frac{\exp \left( 2\omega /T\right) -1}{\exp \left( 2\omega /T(0)\right) -1%
}.  \label{s7}
\end{equation}%
By taking the large $t$ limit with $rt=\tau \,$\ fixed in Eq.~(\ref{s6}),
the following relation holds: 
\begin{equation}
T(\tau )=\frac{\omega }{\ln \left[ 1+\left( \exp \left( \frac{\omega }{T(0)}%
\right) -1\right) \exp \left( -\tau \right) \right] }.  \label{s8}
\end{equation}%
Assuming $\omega \ll T(0)$, we get Eq.~(13). Similarly, using the same
limit, Eq.~(12) is derived by substituting Eq.~(13) into Eq.~(\ref{s7}).

Finally we add a comment regarding another channel, $\Gamma^{\prime}=\Gamma
_{R}\Gamma_{S}$. This yields a different evolution, such that 
\begin{equation*}
\Gamma ^{\prime }\left[ \rho \left( T\right) \right] =\frac{2r}{1+\exp
\left( -\omega /T\right) }|0\rangle \langle 0|+\frac{1-2r+\exp \left(
-\omega /T\right) }{1+\exp \left( -\omega /T\right) }\rho \left( T^{\prime
}\right) .
\end{equation*}%
In this case, imposing $1/T=1/T^{\prime }=0$ yields $s=2r$. Thus, $\Gamma
^{\prime }$ becomes unphysical in the case with $1/2<r\leq 1$ since the
probability $s$ exceeds $1$, which implies that we have noncommutativity of $%
\Gamma _{R}$ and $\Gamma _{S}$. However, if we consider a small $r$ limit in
case ii) to derive the 4-dim BH's thermal properties, the commutator merely
gives $O(r^{2})$ correction terms which do not contribute to the final
results.

\bigskip

\begin{center}
{\large Qubit Fast Scrambling }
\end{center}

Here we show the invariance of the thermal state of a decaying qubit in
Eq.~(9) under fast scrambling and qubit free evolution. We have $N$
three-level identical subsystems.

Let us first comment about the unitary evolution $U$ corresponding to $%
\Gamma $, which describes the dynamics of a single decaying qubit, a
radiation field $\Psi _{R}$, and a soft hair field $\Psi _{S}$. For $%
|\psi\rangle=|+\rangle,\,|-\rangle,\,|0\rangle$, it provides 
\begin{align}
& U|\psi\rangle|\mathrm{vac}\rangle_R|\mathrm{vac}\rangle_S  \notag \\
&=c_+|+\rangle|\mathrm{vac}\rangle_R|\mathrm{vac}\rangle_S
+c_-|-\rangle|\varphi_R^{(\psi)}\rangle_R|\mathrm{vac}\rangle_S+c_0|0\rangle|%
\varphi_R^{(\psi)}\rangle_R|\varphi_S^{(\psi)}\rangle_S,  \label{s100}
\end{align}
where 
\begin{equation*}
|\varphi_R^{(\psi)}\rangle_R\propto 
\begin{cases}
\left(\int \varphi_R(x)\Psi_R^\dag(x) dx\right) |\mathrm{vac}\rangle_R\quad & (\text{%
if }\psi=+) \\ 
\quad|\mathrm{vac}\rangle_R \quad & (\text{if }\psi=-,0)%
\end{cases}%
\end{equation*}
and 
\begin{equation*}
|\varphi_S^{(\psi)}\rangle_S \propto 
\begin{cases}
\left(\int \varphi_S(x)\Psi_S^\dag(x) dx\right) |\mathrm{vac}\rangle_S \quad & (\text{%
if }\psi=+,-) \\ 
|\mathrm{vac}\rangle_S \quad & (\text{if }\psi=0)%
\end{cases}%
\end{equation*}
since the unitary time evolution $U$ conserves the excitation numbers. 
%\begin{eqnarray}
%&&U|\psi \rangle |vac\rangle _{R}|vac\rangle _{S}  \notag \\
%&=&c_{+}|+\rangle |vac\rangle _{R}|vac\rangle _{S}+c_{-}|-\rangle |\varphi
%_{R}\rangle _{R}|vac\rangle _{S}+c_{0}|0\rangle |vac\rangle _{R}|\varphi
%_{S}\rangle _{S},  \label{s100}
%\end{eqnarray}
%where $|\psi \rangle $ is a quantum state of the decaying qubit system, and 
%\begin{eqnarray*}
%c_{-}|\varphi _{R}\rangle _{R} &=&\left( \int \varphi _{R}(x)\Psi _{R}^{\dag
%}(x)dx\right) |vac\rangle _{R}, \\
%c_{0}|\varphi _{S}\rangle _{S} &=&\left( \int \varphi _{S}(x)\Psi _{S}^{\dag
%}(x)dx\right) |vac\rangle _{S}.
%\end{eqnarray}*
Let us assume that the coupling functions $g(x)$ in Eq.~(4) and Eq.~(8) are
almost localized around $x=0$ with large $\left\vert g(0)\right\vert $ as in
the above SM section, and that the wave functions $\varphi _{R}(x)$ and $%
\varphi _{S}(x)$ of the emitted particles already have no overlap with the
support of $g(x)$. This ensures that 
\begin{equation}
g(x)\varphi _{R}(x)=g(x)\varphi _{S}(x)=0.  \label{s101}
\end{equation}%
Then we are able to neglect the correlation of the state in Eq.~(\ref{s100})
in the evaluation of the time evolution of the decaing qubit system due to
the $\left( |+\rangle ,|-\rangle \right) $ sector fast scrambling, which
causes random phase factors $\exp \left( i\delta _{\pm }\right) =$ $c_{\pm
}/\,\left\vert c_{\pm }\right\vert ~$such that 
\begin{equation*}
\overline{c_{+}^{\ast }c_{-}}=\overline{c_{+}^{\ast }}c_{0}=\overline{%
c_{-}^{\ast }}c_{0}=0.
\end{equation*}%
Here each bar stands for the ensemble avarage of the fast scrambling. Theses
properties and Eq.~(\ref{s100}) yield the following decohered state: 
\begin{eqnarray*}
&&\overline{U|\psi \rangle \langle \psi |\otimes |\mathrm{vac}\rangle _{R}\langle
\mathrm{vac}|_{R}\otimes |\mathrm{vac}\rangle _{S}\langle \mathrm{vac}|_{R}U^{\dag }} \\
&=&\overline{\left\vert c_{+}\right\vert ^{2}}|+\rangle \langle +|\otimes
|\mathrm{vac}\rangle _{R}\langle \mathrm{vac}|_{R}\otimes |\mathrm{vac}\rangle _{S}\langle \mathrm{vac}|_{S} \\
&&+\overline{\left\vert c_{-}\right\vert ^{2}}|-\rangle \langle -|\otimes
|\varphi _{R}^{(\psi)}\rangle _{R}\langle \varphi _{R}^{(\psi)}|_{R}\otimes
|\mathrm{vac}\rangle _{S}\langle \mathrm{vac}|_{S} \\
&&+\left\vert c_{0}\right\vert ^{2}|0\rangle \langle 0|\otimes |\varphi
_{R}^{(\psi)}\rangle _{R}\langle \varphi _{R}^{(\psi)}|_{R}\otimes |\varphi
_{S}^{(\psi)}\rangle _{S}\langle \varphi _{S}^{(\psi)}|_{S}.
\end{eqnarray*}%
In the evaluation of the single decaying qubit dynamics, it is possible to
replace $|\varphi _{R}^{(\psi)}\rangle _{R}\langle \varphi
_{R}^{(\psi)}|_{R} $ into $|\mathrm{vac}\rangle _{R}\langle \mathrm{vac}|_{R}$ and $|\varphi
_{S}^{(\psi)}\rangle _{S}\langle \varphi _{S}^{(\psi)}|_{S}$ into $%
|\mathrm{vac}\rangle _{S}\langle \mathrm{vac}|_{S}$ because the emitted particles already
leave the interaction region as seen in Eq.~(\ref{s101}), and the field
quantum states are local vacuum states around the decaying qubit. Hence the
reduced state evolution of the decaying qubit after the fast scrambling can
be described again by using the same channel $\Gamma $ as follows.

\begin{eqnarray*}
&&\mathrm{Tr}_{\Psi _{R}\Psi _{S}}\left[ U\left( \overline{U|\psi \rangle
\langle \psi |\otimes |\mathrm{vac}\rangle _{R}\langle \mathrm{vac}|_{R}\otimes |\mathrm{vac}\rangle
_{S}\langle \mathrm{vac}|_{R}U^{\dag }}\right) U^{\dag }\right] \\
&=&\mathrm{Tr}_{\Psi _{R}\Psi _{S}}\left[ 
\begin{array}{c}
U\left( \overline{\left\vert c_{+}\right\vert ^{2}}|+\rangle \langle +|+%
\overline{\left\vert c_{-}\right\vert ^{2}}|-\rangle \langle -|+\left\vert
c_{0}\right\vert ^{2}|0\rangle \langle 0|\right) \\ 
\otimes |\mathrm{vac}\rangle _{R}\langle \mathrm{vac}|_{R}\otimes |\mathrm{vac}\rangle _{S}\langle
\mathrm{vac}|_{S}U^{\dag }%
\end{array}%
\right] \\
&=&\Gamma \left[ \overline{\left\vert c_{+}\right\vert ^{2}}|+\rangle
\langle +|+\overline{\left\vert c_{-}\right\vert ^{2}}|-\rangle \langle
-|+\left\vert c_{0}\right\vert ^{2}|0\rangle \langle 0|\right] .
\end{eqnarray*}%
This justifies the use of $t$-times successive operations of the channels $%
\Gamma ^{t}\left[ \rho \left( T(0)\right) \right] $ in our time evolution
analysis.

Next let us explain the entanglement structure of the total system. The two
states $|\pm \rangle $ among the three states describe the qubit states.
Assuming that emitted particles do not have spatial overlaps, the
entanglement among decaying qubits and the particles can be computed by use
of an extended model, in which each paricle is treated as a quantum of \ an
independent field. Each channel operation $\Gamma _{R}$ of $\Gamma $
requires a fresh vacuum state $|\mathrm{vac}\rangle $ of a Hawking radiation field $%
\Psi _{R}(x)$. Similarly each $\Gamma _{S}$ of $\Gamma $ \ requires a fresh
vacuum of a soft hair field $\Psi _{S}(x)$. Therefore, $2N$ fields are
required for a one-time operation of $\Gamma ^{\otimes N}$, which means that
the $t$-times iteration of $\Gamma ^{\otimes N}$ requires $2Nt$ fields. For
instance, the original single field $\Psi _{R}$ with $Nt$ particles is
described by a set of $Nt$ different $\Psi _{R}$s with one particle
excitation in each field. The initial state of the fields is a tensor
product of $2Nt$ vacuum states, such that 
\begin{equation*}
\bigotimes_{j=1}^{N}\bigotimes_{k=1}^{t}|\mathrm{vac};R\rangle _{\left( jk\right)
}|\mathrm{vac};S\rangle _{\left( jk\right) },
\end{equation*}%
where the subscripts $(jk)$ discriminate the fields, and $|\mathrm{vac};R\rangle
_{\left( jk\right) }$ ($|\mathrm{vac};S\rangle _{\left( jk\right) }$) is the vacuum
state of the $jk$-th radiation field (soft-hair field).

The total dynamics generated by the full Hamiltonian in Eq.~(11) yields the
following complicated entangled state: 
\begin{align*}
& |\Phi\rangle \\
= & \sum_{s_{1}=\pm}\cdots\sum_{s_{N}=\pm} |s_{1}s_{2}\cdots s_{N}\rangle \\
&\quad \otimes\left(
\sum_{\varphi_{(1,1)}=\mathrm{vac},\phi(1)}\cdots\sum_{\varphi_{(N,t)}=\mathrm{vac},%
\phi(t)}c_{\varphi_{(1,1)},\cdots,\varphi_{(N,t)}}\left( s_{1}s_{2}\cdots
s_{N}\right) \bigotimes_{j=1}^{N}\bigotimes_{k=1}^{t}
|\varphi_{(j,k)};R\rangle_{\left( j k\right) }|\mathrm{vac};S\rangle_{\left( j
k\right) }\right) \\
& +\sum_{j=1}^{N}\sum_{s_{1}=\pm}\cdots\sum_{s_{j-1}=\pm}\sum_{s_{j+1}=\pm
}\cdots\sum_{s_{N}=\pm}|s_{1}s_{2}\cdots0_{j}\cdots s_{N}\rangle \\
&\otimes\left(\sum_{\varphi_{(1,1)}=\mathrm{vac},\phi_1}\cdots\sum_{%
\varphi_{(N,t)}=\mathrm{vac},\phi_t}\sum_{t^{\prime }=1}^t
c_{\varphi_{(1,1)},\cdots,\varphi_{(N,t)}}\left( s_{1}s_{2}\cdots 0_j\cdots
s_{N};t^{\prime }\right) \right. \\
&\left.\quad\quad\bigotimes_{j=1}^{N} \left(\bigotimes_{k=1}^{t}
|\varphi_{(j,k)};R\rangle_{\left( j k\right)
}\right)|\mathrm{vac};S\rangle_{\left(j1\right) }\cdots|\phi(t^{\prime
});S\rangle_{\left(jt^{\prime }\right) }\cdots|\mathrm{vac};S\rangle_{\left(jt\right)
}\right) \\
& +\cdots \\
&
+\sum_{\varphi_{(1,1)}=\mathrm{vac},\phi(1)}\cdots\sum_{\varphi_{(N,t)}=\mathrm{vac},%
\phi(t)}c_{\varphi_{(1,1)},\cdots,\varphi_{(N,t)}}\left(00\cdots
0\right)|00\cdots0\rangle \bigotimes_{j=1}^{N} \bigotimes_{k =1}^{t}
|\varphi_{(j,k)};R\rangle_{\left( j k\right) }|\phi(k) ;S\rangle_{\left( j
k\right) },
\end{align*}
where $|\phi(t);R\rangle$ is a one-particle state of the radiation field
such that 
\begin{equation*}
|\phi (t);R\rangle =\int \phi (x,t)\Psi _{R}^{\dag }(x)dx|\mathrm{vac};R\rangle ,
\end{equation*}%
and $|\phi (t);S\rangle $ is a similar one-particle state of the soft hair
field. Interference among the terms of multi-particle emissions can be
neglected in this model. The entanglement between a decaying qubit and other
systems can be evaluated using $|\Phi \rangle $. The reduced state of the $N$%
-qubit system is 
\begin{align*}
& \mathrm{Tr}_{\Psi _{R},\Psi _{S}}\left[ |\Phi \rangle \langle \Phi |\right]
\\
& =\sum_{s_{1}=\pm }\cdots \sum_{s_{N}=\pm }\sum_{s_{1}^{\prime }=\pm
}\cdots \sum_{s_{N}^{\prime }=\pm }\alpha _{s_{1},\cdots
,s_{N},s_{1}^{\prime },\cdots ,s_{N}^{\prime }}|s_{1}\cdots s_{N}\rangle
\langle s_{1}^{\prime }\cdots s_{N}^{\prime }| \\
& +\sum_{j=1}^{N}\sum_{s_{1}=\pm }\cdots \sum_{s_{j-1}=\pm
}\sum_{s_{j+1}=\pm }\cdots \sum_{s_{N}=\pm }\sum_{s_{1}^{\prime }=\pm
}\cdots \sum_{s_{j-1}^{\prime }=\pm }\sum_{s_{j+1}^{\prime }=\pm }\cdots
\sum_{s_{N}^{\prime }=\pm } \\
& \quad \quad \quad \alpha _{s_{1},\cdots ,0_{j},\cdots ,s_{N},s_{1}^{\prime
},\cdots ,0_{j},\cdots ,s_{N}^{\prime }}|s_{1}\cdots 0_{j}\cdots
s_{N}\rangle \langle s_{1}^{\prime }\cdots 0_{j}\cdots s_{N}^{\prime }| \\
& +\cdots \\
& +\alpha _{0,\cdots 0}|0\cdots 0\rangle \langle 0\cdots 0|
\end{align*}%
for some constants $\{\alpha \}$. Note that the fast scrambling with free
evolution of each qubit makes off-diagonal contributions of $\mathrm{Tr}%
_{\Psi _{R},\Psi _{S}}\left[ |\Phi \rangle \langle \Phi |\right] $ such as $%
|+--\cdots -\rangle \langle -+-\cdots -|$ to vanish when we take a partial
trace and compute a one-qubit reduced state. Thus, the fast scrambling
ensemble average of $\rho _{1}=\mathrm{Tr}_{2,3,\cdots ,N}\mathrm{Tr}_{\Psi
_{R},\Psi _{S}}\left[ |\Phi \rangle \langle \Phi |\right] $ takes the
diagonalized form of%
\begin{equation*}
\overline{\rho _{1}}=(1-p)|0\rangle \langle 0|+p(1-q)|-\rangle \langle
-|+pq|+\rangle \langle +|.
\end{equation*}%
By taking $q=\frac{\exp \left( -\omega /T\right) }{1+\exp \left( -\omega
/T\right) }$, this is rewritten as%
\begin{equation*}
\overline{\rho _{1}}=\left( 1-p\right) |0\rangle \langle 0|+p\rho \left(
T\right) ,
\end{equation*}%
with the temperature $T$ of the decaying qubit. Since the expectation value
of the number of surviving qubits is conserved in the fast scrambling, $p$
is unchanged during this process. Moreover, the conservation of the total
energy ensures that $T$ is equal to the temperature before the scrambling.
Thus, the state of a decaying qubit in Eq.~(9) does not change during the
fast scrambling.

\bigskip

\bigskip

\begin{center}
{\large Average Thermal Entropy of the Total System }
\end{center}

In this section, we evaluate the average thermal entropy of $N$ decaying
qubits. For one decaying qubit, the state is given by $\left( 1-p\right)
|0\rangle \langle 0|+p\rho \left( T\right) $, due to the state typicality.
With probability $p$, a surviving qubit with temperature $T$ is observed.
The entropy is then given by $S_{\mathrm{th}}=-\mathrm{Tr}\left[ \rho \left( T\right)
\ln \rho \left( T\right) \right] $. The vacuum state is observed with
probability $1-p$. Then, no thermal entropy appears. We have \thinspace $N$
identical systems. The probability of finding $n$ surviving qubits is given
by $\left( 
\begin{array}{c}
N \\ 
n%
\end{array}%
\right) p^{n}\left( 1-p\right) ^{N-n}$. Therefore, the average of the total
thermal entropy is computed as 
\begin{equation*}
\sum_{n=1}^{N}\left( 
\begin{array}{c}
N \\ 
n%
\end{array}
\right) p^{n}\left( 1-p\right) ^{N-n}\left( nS_{\mathrm{th}}\right) =NpS_{\mathrm{th}}.
\end{equation*}

\bigskip

\begin{center}
\bigskip{\large Discrepancy between }$S_{\mathrm{BH}}/N$ and $pS_{\mathrm{th}}$
\end{center}

In this section, we derive the large discrepancy between $S_{\mathrm{BH}}/N$ and $%
pS_{\mathrm{th}}$. First of all, $S_{\mathrm{BH}}/N$ is defined using $d\left( S_{\mathrm{BH}}/N\right)
=dE/T$. Due to $E=pE_{\mathrm{th}}$ and $dE_{\mathrm{th}}=TdS_{\mathrm{th}}$, 
\begin{equation}
d\left( S_{\mathrm{BH}}/N\right) =d\left( pS_{\mathrm{th}}\right) -\left( S_{\mathrm{th}}-\frac{E_{\mathrm{th}}}{%
T}\right) dp  \label{s9}
\end{equation}%
holds. The second term on the right-hand side appears due to the time
dependence of $p$, and causes the large deviation of $S_{\mathrm{BH}}/N$ from $%
pS_{\mathrm{th}} $. Let $1/T~$\ be denoted by $\beta $. Using%
\begin{equation*}
S_{\mathrm{th}}-\frac{E_{\mathrm{th}}}{T}=\ln \left( 1+\exp \left( -\beta \omega \right)
\right) +\frac{\beta \omega \exp \left( -\beta \omega \right) }{1+\exp
\left( -\beta \omega \right) }-\frac{\beta \omega \exp \left( -\beta \omega
\right) }{1+\exp \left( -\beta \omega \right) }=\ln \left( 1+\exp \left(
-\beta \omega \right) \right) ,
\end{equation*}%
and 
\begin{equation*}
\beta =p\beta \left( 0\right)
\end{equation*}%
assuming $\omega \ll T(0)$, the integration of Eq.~(\ref{s9}) provides 
\begin{equation}
S_{\mathrm{BH}}/N=p\left( S_{\mathrm{th}}-\ln 2\right) +\int_{0}^{p}dp^{\prime
}\int_{0}^{p^{\prime }\beta (0)}E_{\mathrm{th}}\left( \beta ^{\prime }\right) d\beta
^{\prime }.  \label{s10}
\end{equation}%
By taking a small $\beta $ in Eq.~(\ref{s10}), the term of $O\left(
T(0)/T\right) $ in the right-hand side vanishes because $S_{\mathrm{th}}$ tends to $%
\ln 2$. Thus, the relation $S_{\mathrm{BH}}/N=O\left( T(0)^{2}/T^{2}\right) $ is
reproduced, even though $pS_{\mathrm{th}}=$ $O\left( T(0)/T\right) $.


\bigskip

\begin{thebibliography}{99}
\bibitem{RT} S. Ryu and T. Takayanagi, Phys. Rev. Lett. {\bf 96}, 181602 (2006).

\bibitem{J} T. Jacobson, Phys. Rev. Lett. {\bf 116}, 201101 (2016).

\bibitem{KP} A. Kitaev and J. Preskill, Phys. Rev. Lett. {\bf 96}, 110404 (2006).

\bibitem{LW} M. Levin and X.-G. Wen, Phys. Rev. Lett. {\bf 96}, 110405 (2006).

\bibitem{exp} R. Islam, R. Ma, P. M. Preiss, M. E. Tai, A. Lukin, M.
Rispoli, and M. Greiner, Nature (London) {\bf 528}, 77 (2015).

\bibitem{h} S. W. Hawking, Commun. Math. Phys. {\bf 43}, 199 (1975).

\bibitem{h2} S. W. Hawking, Phys. Rev. D {\bf 72}, 084013 (2005).

\bibitem{page} D. N. Page, Phys. Rev. Lett. {\bf 71}, 3743 (1993).

\bibitem{HSS} M. Hotta, K. Sasaki, and T. Sasaki, Classical Quantum
367 Gravity {\bf 18},
1823 (2001).

\bibitem{HPS} S. W. Hawking, M. J. Perry, and A. Strominger, Phys. Rev.
Lett. {\bf 116}, 231301 (2016).

\bibitem{HTY} M. Hotta, J. Trevison, and K. Yamaguchi, Phys. Rev. D {\bf 94}, 083001 (2016).

\bibitem{HPS2} S. W. Hawking, M. J. Perry, and A. Strominger,  J. High
373 Energy Phys. 05 (2017) 161.

\bibitem{hanada} E. Berkowitz, M. Hanada, and J. Maltz, Phys. Rev. D {\bf 94},
126009 (2016).

\bibitem{M} S. D. Mathur, Classical Quantum Gravity {\bf 26}, 224001 (2009).

\bibitem{G} S. B. Giddings, Phys. Rev. D {\bf 85}, 124063 (2012).

\bibitem{A} S. G. Avery,  J. High Energy Phys. 01 (2013) 176.

\bibitem{L} E. Lubkin, J. Math. Phys. (N.Y.) {\bf 19}, 1028 (1978).

\bibitem{Seth} S. Lloyd and H. Pagels, Ann. Phys. (N.Y.) {\bf 188}, 186 (1988).

\bibitem{sugita} A. Sugita, RIMS Kokyuroku (Kyoto) {\bf 1507}, 147 (2006).

\bibitem{sugita2} A. Sugita, Nonlinear Phenom. Complex Syst. {\bf 10}, 192 (2007).

\bibitem{fs} Y. Sekino and L. Susskind, , J. High Energy Phys. 10 (2008) 065.

\bibitem{SM} See Supplemental Material for more detailed derivations.

\bibitem{B} S. L. Braunstein, S. Pirandola, and K. \.{Z}yczkowski, Phys.
Rev. Lett. {\bf 110}, 101301 (2013).

\bibitem{AMPS} A. Almheiri, D. Marolf, J. Polchinski, and J. Sully, J. High
Energy Phys. 02 (2013) 062.

\bibitem{HS} M. Hotta and A. Sugita, Prog. Theor. Exp. Phys. {\bf 2015}, 123B04 (2015).
\end{thebibliography}
\end{document}